# Concerted Electron-Ion Transport by Polyacrylonitrile Elucidated with Reactive Deep Learning Potentials

Rajni Chahal-Crockett†*, Michael D. Toomey, Logan T. Kearney*, Yawei Gao, Joshua T. Damron, Amit K. Naskar, Santanu Roy*

Chemical Science Division, Oak Ridge National Laboratory, Oak Ridge, TN-37830, United States

Email: rchahal@tntech.edu; kearneylt@ornl.gov; roys@ornl.gov

**ABSTRACT:** Charge transport in polymers, such as Polyacrylonitrile (PAN), is crucial for electronics and energy storage. For instance, PAN can transport cations e.g., $Li^+$, by facilitating dynamic cation-nitrile coordination in batteries. However, little is known regarding the underlying role of complex reactive polymer configurations. Herein, we develop a deep-learning potential, trained on *ab initio* energies and forces of nonequilibrium reactive PAN configurations, to unravel the kinetics of PAN cyclization initiated by a nucleophile ($OH^-$ dissociated from LiOH) attacking the terminal nitrile carbon. We find, based on the reaction free-energetics, rates, and charge analysis, that the nucleophile attack producing the first ring is the rate-limiting step, which subsequently triggers $Li^+$-coupled electron transfer along the PAN backbone, causing ~$10^4$ times faster sequential ring-formation of the remaining nitriles. PAN's extended configurations, where dipolar and H-bonding interactions are minimal, enable such rapid kinetics. By validating our computational findings with IR and NMR experiments, we establish a pathway for designing reactive polymers with enhanced charge transport for energy applications.

Charge transport, a fundamental event in condensed-phase systems, can be triggered via local molecular interactions,[1,2,3] medium reorganization,[4,5,6,7] and impurities or defects.[8,9,10,11,12] Among many materials or molecules that exhibit charge transport events, polymers are particularly critical due to their broad range of applications, from organic electronics to energy storage. For example, redox-active polymers such as polyacetylene and polypyrrole are used to transport electrons through battery electrodes,[13,14,15,16] while frequently-used ion-conducting polymer electrolytes in batteries are polyethylene oxide and polyacrylonitrile (PAN).[17,18,19,20,21,22,23] However, a major bottleneck for efficient charge transport by such polymers may arise from their complex diverse configurations. Several studies[24,25,26,27] showed that more ordered and extended chain conformations can lead to faster electron transport due to enhanced electronic coupling between repeating units. Other studies[28,29,30,31,32] indicate that amorphous, flexible polymer matrix can enhance ion diffusion. Nevertheless, the mechanism by which tunable polymer configurations can enable energy-efficient coupled electron-ion transport, critical for reaction-driven energy-storage systems,[33,34,35,36] remains poorly understood.

Recently, the remarkable success[37,38,39,40,41,42,43,44,45,46] in integrating deep-learning methods, electronic structure theory, and molecular dynamics (MD) simulations demonstrates the feasibility of accessing reactive chemistry, while exploring the large configurational space of a polymer. For instance, energies and forces obtained from density functional theory (DFT)-based *ab initio* MD of oligomers and their self-assembly can be leveraged to train a neural-network interatomic potential (NNIP).[47] Subsequently, NNIP-based MD—referred to as NNMD—when combined with enhanced sampling techniques, such as umbrella sampling or metadynamics,[48,49,50,51] provides potentially a powerful approach for investigating a wide range of reactive events, from local catalytic effects to cascades of bond-forming reactions in polymers. Herein, we demonstrate that an NNIP, trained on the DFT data from nonequilibrium reactive configurations enhanced-sampled along cyclization reaction paths of PAN, enables NNMD to examine the process of PAN-mediated electron-coupled ion transport.

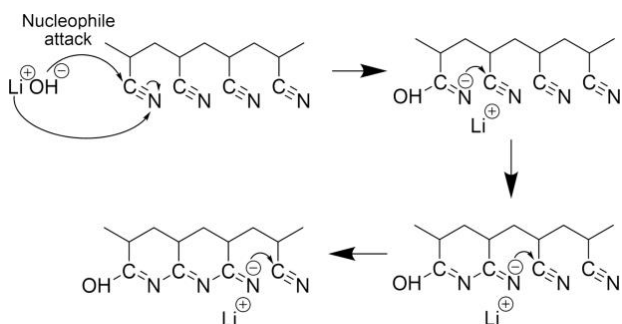

**Figure 1**. The mechanism of cyclization in a 4-unit PAN chain, initiated by the $OH^-$ nucleophile. As the cyclization progresses, $Li^+$ follows the electron localized on N, which again attacks the next -C≡N to continue the cyclization process.

PAN, as aforementioned, is a promising component of battery electrolytes with superior mechanical properties and electrochemical stability.[52,22,53,23,54] It can leverage its nitrile (−C≡N) groups as coordination sites for cations, e.g., $Li^+$ ions, guiding their transport within a battery through the polymer matrix. Herein, *both* computationally and experimentally, we showcase that PAN can be cyclized by an $OH^-$ nucleophile attack at room temperature, which otherwise requires 200-300°C.[55,56,57] Such a cyclization process enables electron-coupled $Li^+$ transport. As depicted in **Figure 1**, $OH^-$ attacks C of a terminal −C≡N group, converting it to a −C=N$^{(-)}$ (excess electron-localization at N) group, which again, as a nucleophile, attacks C of the adjacent −C≡N group, causing sequential cyclization up to the last unit and formation of a ladderlike configuration. Interestingly, Li$^+$ moves along the cyclization steps, coordinating with the −C=N$^{(-)}$ group, i.e., follows the propagation of the electron along the PAN backbone. By exploring reaction free-energy landscapes and employing transition state theory, we find that such electron-coupled Li$^+$ transfer can be effectively instantaneous in extended PAN configurations, provided the initial nucleophilic attack has already initialized. Our findings were corroborated through time dependent spectroscopic measurements of the PAN-LiOH system dissolved in polar solvents.

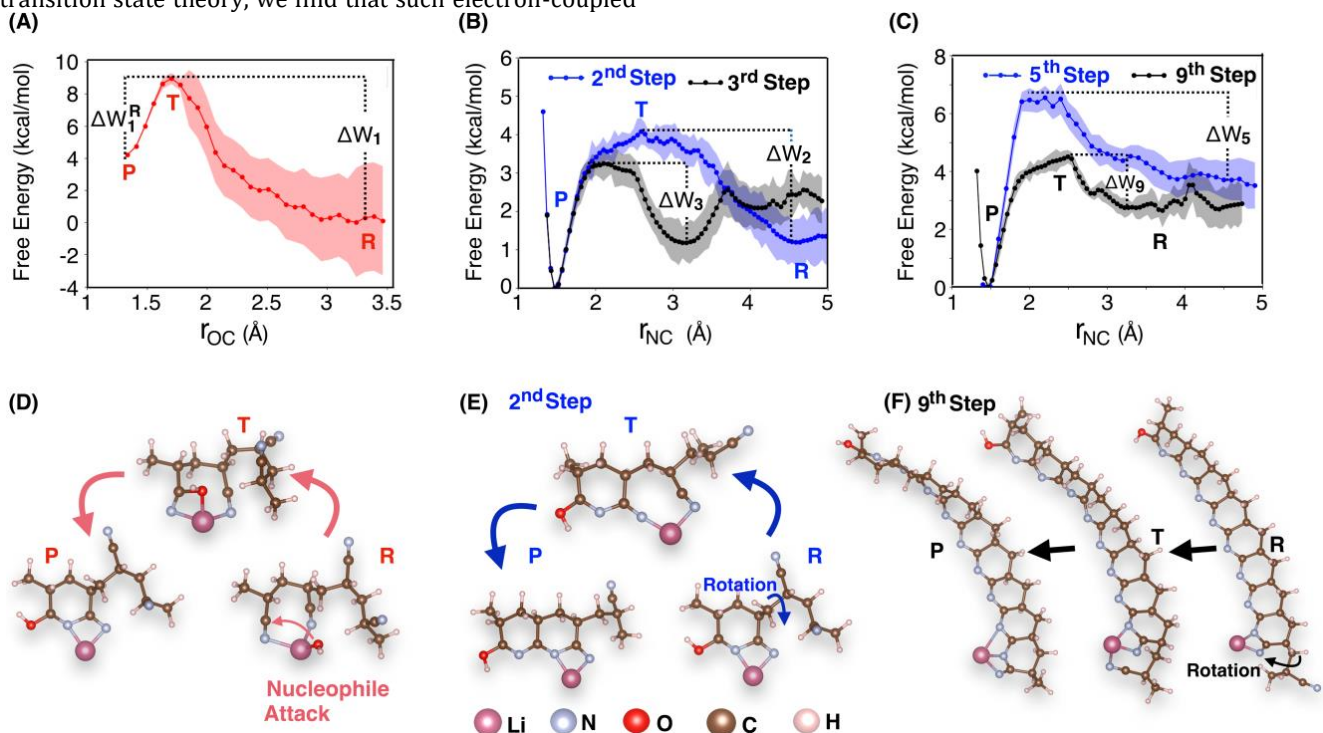

**Figure 2.** For a 4-mer PAN, the free-energy profiles for the nucleophile (OH$^-$) attack (A) and the second and third cyclization steps (B), highlighting the associated barriers, $\Delta W_1$, $\Delta W_2$, and $\Delta W_3$. The same for a 10-mer PAN for the fifth and nineth steps with the barriers, $\Delta W_4$ and $\Delta W_9$ (C). Snapshots illustrating structural changes in going from the reactant (R) to transition (T) to product (P) states of the initial nucleophile attack and second cyclization steps in a 4-mer PAN (D, E) and of the nineth cyclization step of a 10-mer PAN (F).

Towards achieving quantitative details of the reaction free-energies and kinetics and the underlying configurational effects, we explore 8 generations (Gens 0-7) of NNIPs. Gen 0, developed in our earlier study,[47] is suitable for simulating nonreactive PAN chains and bulk systems. As detailed in **Supporting Information (SI),** Gen 1 evolves to Gen 7, considering different reactions and configurational effects (**Figures S1, S2**). The main consideration is to account for energies and forces from the AIMD simulations (detailed in **SI**) of the PAN configurations generated in the umbrella sampling of two reaction coordinates; the distance between the hydroxyl oxygen and C of the terminal −C≡N group ($r_{OC}$) and the distance between N of a −C=N$^{(-)}$ group and C of the adjacent −C≡N group ($r_{CN}$) (**Figure 1**). The inclusion of other critical effects such as chain-chain dipolar and H-bonding interactions (considering uncyclized and partially cyclized chains) in dimeric and bulk configurations in the training ultimately leads to a robust, accurate, and stable NNIP, Gen 7 (**Figures S3-S11**), which is used for all final computations of reaction free-energies.

**Figure 2** depicts the free-energy profiles of the cyclization steps for a 4-mer and a 10-mer chains in the gas-phase environment. We choose these oligomers to mimic local structures of an extended large polymer, avoiding astronomically-high configurational barriers that may arise from strong −C≡N dipole-dipole interactions and −C≡N⋯ ·HC− hydrogen bonding in folded, contracted, or coiled bulky structures (**Figure S14**) typically expected in the gas-phase environment. Naturally, the free-energy profiles in **Figure 2** predominantly reveal the barriers for a combined effect of bond rotation, moderate configurational change, and new bond breaking or forming through the variation of $r_{OC}$ and $r_{CN}$. As OH$^-$ attacks C of a terminal −C≡N (i.e., $r_{OC}$ reduces) of a 4-mer PAN from a close distance (~3.5 Å) (**Figures 2A, 2D**), it dissociates from Li$^+$ and helps form the first ring by overcoming a barrier ($\Delta W_1$) of ~9 kcal/mol. This is higher than the barrier for the reverse process ($\Delta W_1^R$ ~5 kcal/mol), which is attributed to the gas-phase environment that makes LiOH dissociation, and consequently, the OH$^-$ attack on C less probable. However, once the first ring has formed, the barriers for forming the remaining rings ($\Delta W_2$ and $\Delta W_3$) are within 3 kcal/mol, as indicated by the free-energy profiles of $r_{CN}$ (**Figure 2B**). For the 10-mer PAN, the cyclization barriers for the fifth and ninth units ($\Delta W_5$ and $\Delta W_9$ in **Figure 2C**) are similarly small (≤3 kcal/mol). Interestingly, regardless of whether it's a 4-mer or a 10-mer chain, the cyclization barrier decreases for each successive unit, suggesting a sequential reduction in the configurational disorder as cyclization advances towards the final unit. Furthermore, the energy barrier for each intermediate cyclization step is lower than that for its reversal, indicating a higher probability of successful sequential cyclization to the final unit once the process begins. An interesting structural change for an intermediate cyclization step (**Figures 2E, 2F**) as $r_{CN}$ varies is the C-C bond rotation that aligns the $-C\equiv N$ and $-C=N^{(-)}$ groups on the same side, positioning them close to each other at the transition state and facilitating ring formation.

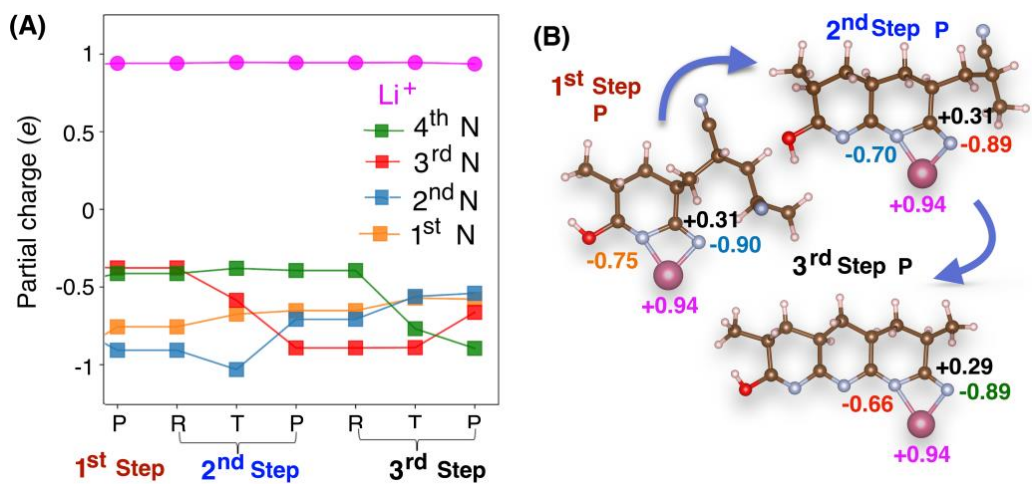

**Figure 3**. Partial charge of the N atoms in a 4-mer PAN in the R, T, and P states of all three cyclization steps (A). A quadrupole-like configuration is formed in the product state of each step (B).

To quantify the reaction kinetics, particularly to estimate how fast an intermediate cyclization step is, compared to the initialization step, we employ TST within the harmonic approximation: $k = \left(\frac{k_B T}{h}\right)\exp[-\Delta W/k_B T]$ where $k$ is the reaction rate for the temperature T and the barrier $\Delta W$. $k_B$ and $h$ are respectively the Boltzmann and Planck constants. Given the initial barrier, ~9 kcal/mol and a maximum barrier, ~3 kcal/mol for any intermediate step at 300 K, the initialization occurs at a rate of ~1.7 $\mu s^{-1}$, while an intermediate step occurs at a rate of 0.0403 $ps^{-1}$ or faster. That is, the latter is four orders of magnitude faster. The reader is reminded that sequential cyclization occurs due to the successive attack of the electron localized on N of a $-C=N^{(-)}$ unit onto C of the succeeding $-C\equiv N$ unit. Therefore, electron transfer through the N atoms during cyclization and the paired transfer of Li+ with the electron—*both* should occur on the same fast timescale along an extended polymer configuration.

To provide evidence of electron-coupled Li+ transfer during the cyclization process, we perform natural population analysis[58] on the NNMD snapshots of a 4-mer PAN, representing R, P, and T states. Using a variety of electronic structure methods, we find very similar partial atomic charges across these methods (see **SI**). **Figure 3A** shows, from the MP2-level[59] calculations, how the charge on each N of the 1st to 4th PAN units changes as the cyclization progresses to the 4th unit. In specific, the negative charge on the 4th N is maintained at ~-0.5e during the 1st and 2nd cyclization steps, but it increases to ~-0.9e after the 3rd cyclization step. In reverse, the ~-0.9e charge on the 2nd N of the singly-cyclized unit reduces to ~-0.5e after all three units are cyclized. **Figure 3B** highlights the charges on the N, Li+, and nearby C atoms in the product-state snapshots of all three steps. The -0.9e charge confirms the electron localization, i.e., the $-C=N^{(-)}$ group formation, which propagates from one product state to another as the cyclization continues. Interestingly, considering the charges on Li+ and its nearest N and C atoms, it appears that a quadrupole-like local configuration, $Li^{(+)}N^{(-)}C^{(+)}N^{(-)}$, moves from one cyclized state to another. Going back to **Figure 2F**, we find that the transition state has a similar configuration, where the dual effects of the nucleophile $N^{(-)}$ attacking $C^{(+)}$ and the electrophile $Li^{(+)}$ attacking $N^{(-)}$ are apparent, and are likely what is causing rapid cyclization and charge transfer at any intermediate step.

We recognize that, for any PAN chain-length, a fully-extended configuration can make joint nucleophile-electrophile attacks and consequent cyclization most efficient due to minimal configurational barriers. **Figures S15-S17** justifies this—only a fully-extended configuration of a 20-mer PAN, starting with the 1st-ring cyclized, exhibit nearly-complete cyclization and Li+ transport to the last unit during a 400 picosecond-long unbiased NNMD, while all other configurations remain stuck in the partially-cyclized state. In practice, extended configurations can be accessed by dissolving PAN in polar solvents such as DMF and DMSO, where $-C\equiv N\cdots HC-$ hydrogen bonds can break rapidly, allowing for efficient reactions (**Figures S18**). Therefore, we validate our theoretical findings by measuring IR and 1H NMR spectra of chain-extended PAN dissolved in a LiOH containing DMSO solution (See **SI**). **Figure 4** summarizes key findings, highlighting effective room-temperature cyclization in the solution state. The increased conjugation is supported by IR analysis (**Figure 4A**), which shows a progressive decrease in the symmetric $-C\equiv N$ stretching band (~2240 $cm^{-1}$) with time, accompanied by the emergence of aromatic (~1600 $cm^{-1}$) and imine (~1660 $cm^{-1}$) spectral features. Notably, these positions are slightly blue shifted in comparison to thermally induced cyclization[60] which we attribute to the presence of the charged complexes. The DFT-based IR spectra calculations corroborate these findings with the appearance of the

–C=N$^{(-)}$ stretching vibration peak at ~1665 cm$^{-1}$ (**Figure S20**).

To better quantify the kinetics, we formulated a LiOH/PAN mixture in d6-DMSO and continuously collected $^1$H NMR spectra over the course of several days. The decay of the characteristic methylene groups in the PAN backbone, which are consumed during cyclization and dehydrogenation,[61] is plotted as a function of time in **Figure 4B**. A good fit was attained with a biexponential function, suggesting two kinetically distinct phases in the Li-promoted cyclization. This behavior is consistent with ionic initiation, in which rapid cyclization occurs initially in more accessible or pre-aligned nitrile sequences, followed by a slower phase as reactions propagate into less mobile regions of the polymer where conformational rearrangement and reduced chain mobility limit the rate of further ladder formation.

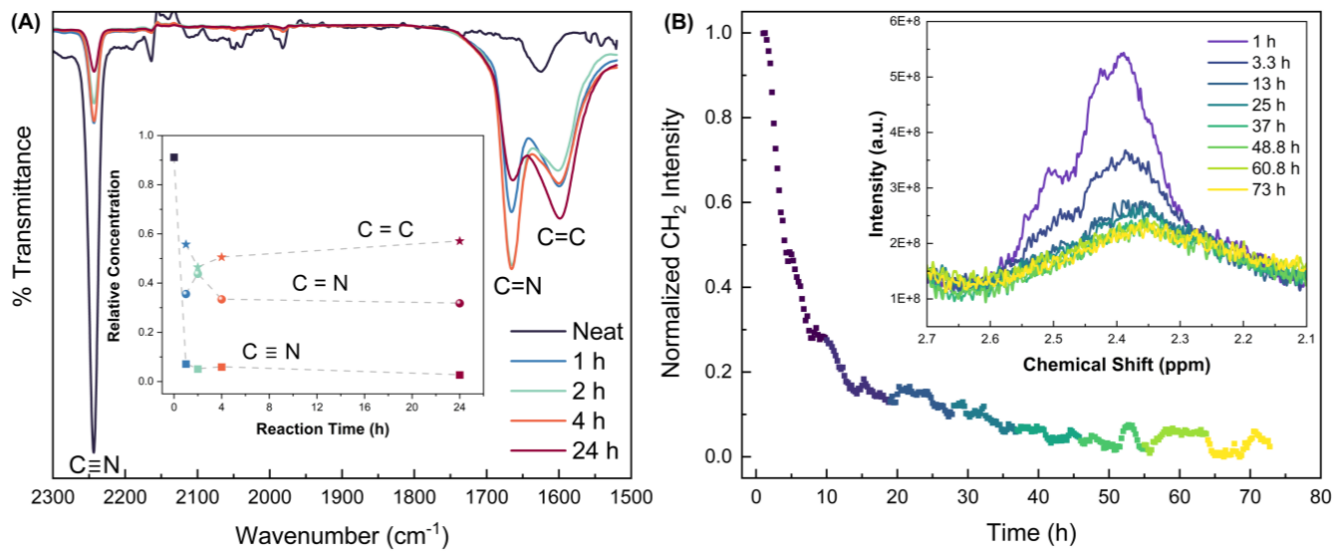

**Figure 4**. IR spectra of a PAN cyclization products following exposure to LiOH in solution for varying durations and relative fractions of resulting peaks (inset) (A). Decay rate of methylene peak area during in-situ $^1$H NMR, indicative of nitrile group consumption. Spectral sampling during the experiment (inset) (B).

The computational investigation of the cyclization reactions in solvent mixtures is subject to future research. However, the mechanism of electron-coupled Li$^+$ ion transport, particularly the propagation of the quadrupole-like local configuration, Li$^{(+)}$N$^{(-)}$C$^{(+)}$N$^{(-)}$, along the polymer backbone, should persist in the solvent, because the Li$^+$-solvent interaction is relatively weak, allowing Li$^+$ to follow the electron. The current study suffices to establish a deep-learning-enabled computational platform for studying reactivity and charge transfer in polymers. Furthermore, it will guide synthetic chemists to design polymeric materials for efficient charge conductions in the energy storage systems.

## ASSOCIATED CONTENT

**Data Availability Statement**

NNIP potential will be available at a GitHub repository.

**Supporting Information**

The Supporting Information is available free of charge on the ACS Publications website.
Comprehensive Details of Computational Methods; Experimental Procedure; Supporting Figures S1-S21; Supporting Tables S1-S12

## AUTHOR INFORMATION

**Corresponding Author**

Rajni Chahal-Crockett – Chemical Sciences Division, Oak Ridge National Laboratory, Oak Ridge, Tennessee 37830, USA
Email: rchahal@tntech.edu

Logan T. Kearney - Chemical Sciences Division, Oak Ridge National Laboratory, Oak Ridge, Tennessee 37830, USA
Email: kearneylt@ornl.gov

Santanu Roy - Chemical Sciences Division, Oak Ridge National Laboratory, Oak Ridge, Tennessee 37830, USA
Email: roys@ornl.gov

**Present Addresses**

†Mechanical and Nuclear Engineering, Tennessee Technological University, Cookeville, TN 38501

**Author Contributions**

The manuscript was written through contributions of all authors. All authors have given approval to the final version of the manuscript.

**Notes**

The authors declare no competing financial interest. This manuscript has been authored by UT-Battelle, LLC under Contract No. DE-AC05−00OR22725 with the U.S. Department of Energy. The United States Government retains and the publisher, by accepting the article for publication, acknowledges that the United States Government retains a nonexclusive, paid-up, irrevocable, worldwide license to publish or reproduce the published form of this manuscript, or allow others to do so, for United States Government purposes. The Department of Energy will provide public access to these results of federally sponsored research in accordance with the DOE Public Access Plan (http://energy.gov/ downloads/doe-public-access-plan).

## ACKNOWLEDGMENT

This research was sponsored by the U.S. Department of Energy (DOE)'s Office of Critical Minerals and Energy Innovation, Hydrogen and Fuel Cell Technologies Office Program (Award DE-LC-000L83). This research used resources of the Oak Ridge Leadership Computing Facility at the Oak Ridge National Laboratory, which is supported by the Office of Science of the U.S. Department of Energy under Contract No. DE-AC05000OR22725. Additionally, this research used resources of the National Energy Research Scientific Computing Center, a DOE Office of Science User Facility supported by the Office of Science of the U.S. Department of Energy under Contract No. DE-AC02-05CH11231 using NERSC award BES-ERCAP0033005. M.D.T., L.T.K., Y.G., J.T.D. (spectroscopic validation work) acknowledge support from DOE's Office of Science, Basic Energy Sciences, Materials Sciences and Engineering Division (FWP #ERKCK60], under Contract DE-AC05-00OR22725 with UT-Battelle, LLC. We thank Dr. Sumit Gupta for helpful discussions.

## Table of Contents Artwork

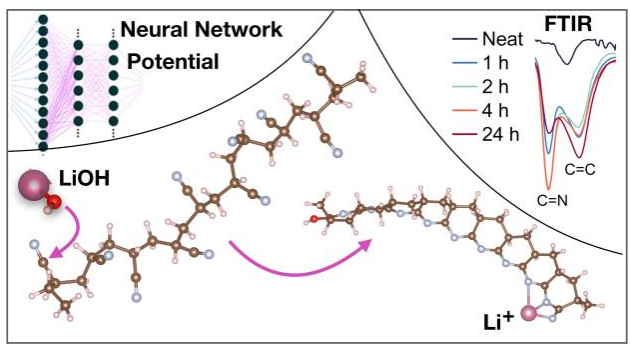